\documentclass[conference]{IEEEtran}
\usepackage{cite}
\usepackage{amsmath,amssymb,amsfonts}
\usepackage{algorithmic}
\usepackage{graphicx}
\usepackage{textcomp}
\usepackage{xcolor}
\usepackage[hyphens]{url}
\usepackage{braket}
\usepackage[linesnumbered,ruled,vlined]{algorithm2e}
\usepackage{subcaption}

\def\BibTeX{{\rm B\kern-.05em{\sc i\kern-.025em b}\kern-.08em
    T\kern-.1667em\lower.7ex\hbox{E}\kern-.125emX}}

\title{TensorQC: Towards Scalable Distributed Quantum Computing via Tensor Networks}
\author{
  \IEEEauthorblockN{Wei Tang\textsuperscript{1}}
  \IEEEauthorblockA{Department of Computer Science\\
  Princeton University\\
  Princeton, NJ, USA\\
  weit@alumni.princeton.edu}
  \and
  \IEEEauthorblockN{Margaret Martonosi}
  \IEEEauthorblockA{Department of Computer Science\\
  Princeton University\\
  Princeton, NJ, USA\\
  mrm@princeton.edu}
}

\begin{document}
\maketitle
\thispagestyle{plain}
\pagestyle{plain}


\begin{abstract}

  A quantum processing unit (QPU) must contain a large number of high quality qubits to produce accurate results for problems at useful scales.
  In contrast, most scientific and industry classical computation workloads happen in parallel on distributed systems,
  which rely on copying data across multiple cores.
  Unfortunately, copying quantum data is theoretically prohibited due to the quantum non-cloning theory.
  Instead, quantum circuit cutting techniques cut a large quantum circuit into multiple smaller subcircuits,
  distribute the subcircuits on parallel QPUs and reconstruct the results with classical computing.
  Such techniques make distributed hybrid quantum computing (DHQC) a possibility but also introduce an exponential classical co-processing cost in the number of cuts and easily become intractable.
  This paper presents TensorQC, which leverages classical tensor networks
  to bring an exponential runtime advantage over state-of-the-art parallelization post-processing techniques.
  As a result, this paper demonstrates running benchmarks that are otherwise intractable for a standalone QPU and prior circuit cutting techniques.
  Specifically, this paper runs six realistic benchmarks using QPUs available nowadays and a single GPU,
  and reduces the QPU size and quality requirements by more than $10\times$ over purely quantum platforms.

\end{abstract}
\footnotetext[1]{This work was done prior to joining Amazon.}
\section{Introduction}

Quantum Computing (QC) presents itself as a promising alternative to traditional classical computing.
Various quantum algorithms offer potential runtime benefits compared to the best-known classical counterparts,
notably in unstructured database searches~\cite{grover1996fast},
optimization tasks~\cite{moll2018quantum} and integer factorization~\cite{shor1999polynomial}.
In practice, researchers design a quantum algorithm as an $n$-qubit quantum circuit.
A single QPU then repeatedly executes this circuit,
measuring its output as binary strings.
The output binary strings then manifest as a binary state probability distribution,
where solution states are distinguished by significantly higher quantum state amplitudes compared to non-solution states.
Accurate QPU execution and measurement are critical,
requiring a sufficient number of trials.

However, QPUs must satisfy highly demanding size and quality requirements to reliably run useful benchmarks.
First, a QPU must have at least $n$ qubits to even run an $n$-qubit circuit at all.
Furthermore, these $n$ qubits must be accurate and robust enough to support the quantum workload without accumulating too much noise to produce quality solutions.
Unfortunately, these two requirements put a heavy toll on the hardware.
As an example, the famous Shor's integer factorization algorithm requires millions of physical qubits to construct enough logical qubits to run problems at practical scales~\cite{suchara2013comparing}.

Currently, classical scientific and industrial workloads predominantly operate in parallel,
moving away from dependency on a single, powerful core.
Classical parallel computing harnesses data and model parallelism~\cite{shoeybi2019megatron,huang2019gpipe},
involving the distribution of classical data across multiple computing units.
For example, classical simulations of quantum circuits and systems usually use large supercomputers~\cite{liu2021closing} with millions of cores.
Similarly, numerous machine learning applications, like AlphaFold~\cite{jumper2021highly},
leverage hundreds of GPUs to manage complex computations efficiently.

Although the concept of divide and conquer is straightforward,
QC introduces distinct challenges.
The quantum no-cloning theorem~\cite{wootters1982single} prohibits the duplication of quantum data,
rendering traditional classical strategies like divide and conquer inapplicable to QC.
However, the development of quantum circuit cutting theory~\cite{peng2020simulating}
and its practical implementations~\cite{tang2021cutqc} have paved the way for distributing large quantum circuits across multiple QPUs.
This method relies on classical co-processing to handle the division effectively without copying quantum data.

Circuit cutting involves three stages.
First, circuit cutting cuts a large quantum circuit into several smaller subcircuits.
A distributed set of less powerful QPUs then run the subcircuits in parallel.
Classical co-processing then reconstructs the complete output from these small subcircuits without ever copying quantum data across QPUs.
This is analogous to classical parallel computing,
which distributes a large workload among less powerful computing units at the expense of communication and reconstruction costs.
Section~\ref{sec:background} explains in more details.

Dividing a large quantum circuit into smaller subcircuits offers significant benefits compared to using a single, large QPU.
Firstly, these subcircuits require fewer qubits, which alleviates the stringent size constraints on QPUs.
For instance, some algorithms demonstrated the capability to operate with fewer than half the number of qubits typically necessary~\cite{tang2021cutqc}.
Secondly, the reduced number of quantum gates in subcircuits lessens the demand for high qubit quality,
enhancing the accuracy of the outcomes.
The simpler, shallower architecture of subcircuits results in reduced crosstalk~\cite{mundada2019suppression},
decoherence~\cite{klimov2018fluctuations},
gate errors~\cite{arute2019quantum} and control challenges~\cite{abdelhafez2020universal}.
Overall, quantum circuit cutting leverages both modest QPUs and classical computing resources to facilitate DHQC effectively.

Hybrid and distributed architectures are poised to become the standard framework for future quantum computing,
thanks to their promising advantages.
Indeed, IBM has announced plans to base its quantum systems on these techniques,
considering circuit cutting a cornerstone of practical quantum computing~\cite{bravyi2022future}.
Despite the commercial interest and the encouraging results from preliminary works,
several fundamental challenges limit the scalability of quantum circuit cutting:
\begin{enumerate}
   \item The classical co-processing overhead increases exponentially with the number of cuts to split a quantum circuit.
   Realistic benchmarks require multiple cuts and the computational overhead quickly becomes intractable.

   \item Classically reconstructing the full binary state probability distribution scales exponentially with the number of qubits as $2^n$.
   \item Identifying cuts for arbitrary quantum circuits to reduce the hybrid computation cost while being subject to hardware constraints
   reduces to an NP-complete graph partition problem~\cite{karypis1998fast}.
   
\end{enumerate}

This paper presents \emph{tensor network contraction},
\emph{heavy state selection},
and \emph{constrained graph partitioning heuristics}
to tackle these challenges.
We ran several realistic quantum benchmark algorithms such as QAOA~\cite{saleem2020approaches},
AQFT~\cite{barenco1996approximate},
Supremacy~\cite{arute2019quantum}, W-state~\cite{dur2000three, diker2016deterministic}, and GHZ~\cite{greenberger1989going} to evaluate the performance of TensorQC.
Our contributions include the following:
\begin{enumerate}
   \item TensorQC paves the way towards scalable distributed QC.
   Our experiments run quantum circuits of up to $200$ qubits using QPUs available nowadays and a single GPU.
   These benchmarks are otherwise intractable using classical simulations, single-QPU or even existing circuit cutting techniques.
   \item TensorQC pioneers the development of versatile distributed quantum architectures.
   This paper showcases a diverse array of benchmarks,
   including circuits characterized by high connectivity and entanglement,
   highlighting the broad applicability and robustness of the technology.
   \item TensorQC unlocks new possibilities for executing large-scale quantum applications.
   This paper integrates cutting-edge classical computing methods from tensor networks with medium-scale QPUs.
   This fusion demonstrates that circuit cutting techniques represent a promising approach to scale up quantum computation systems.
\end{enumerate}

The rest of the paper is organized as follows:
Section~\ref{sec:background} explains the circuit cutting process and its challenges.
Section~\ref{sec:co_processing} introduces tensor networks,
establishes using tensor network contractions for DHQC co-processing,
and proves its exponential runtime advantage over prior parallelization techniques.
Section~\ref{sec:heavy_state_selection} develops an automatic state selection protocol to only retain the significant binary states.
Section~\ref{sec:cost_model} designs a hybrid cost model to capture the combined quantum and classical runtime costs to execute a quantum circuit.
Section~\ref{sec:locate_cuts} presents graph partition heuristics to identify high-quality cuts quickly based on the hybrid cost model.
Section~\ref{sec:methodology} sets up the experiment methodology, and Section~\ref{sec:results} presents the results.
Lastly, Sections~\ref{sec:related_work},~\ref{sec:conclusion} present related works and conclusions.
\section{Background}\label{sec:background}
This section introduces the quantum circuit cutting theory and identifies its key challenges.

\subsection{Circuit Cutting Theory}
\begin{figure*}[t]
    \centering
    \includegraphics[width=\linewidth]{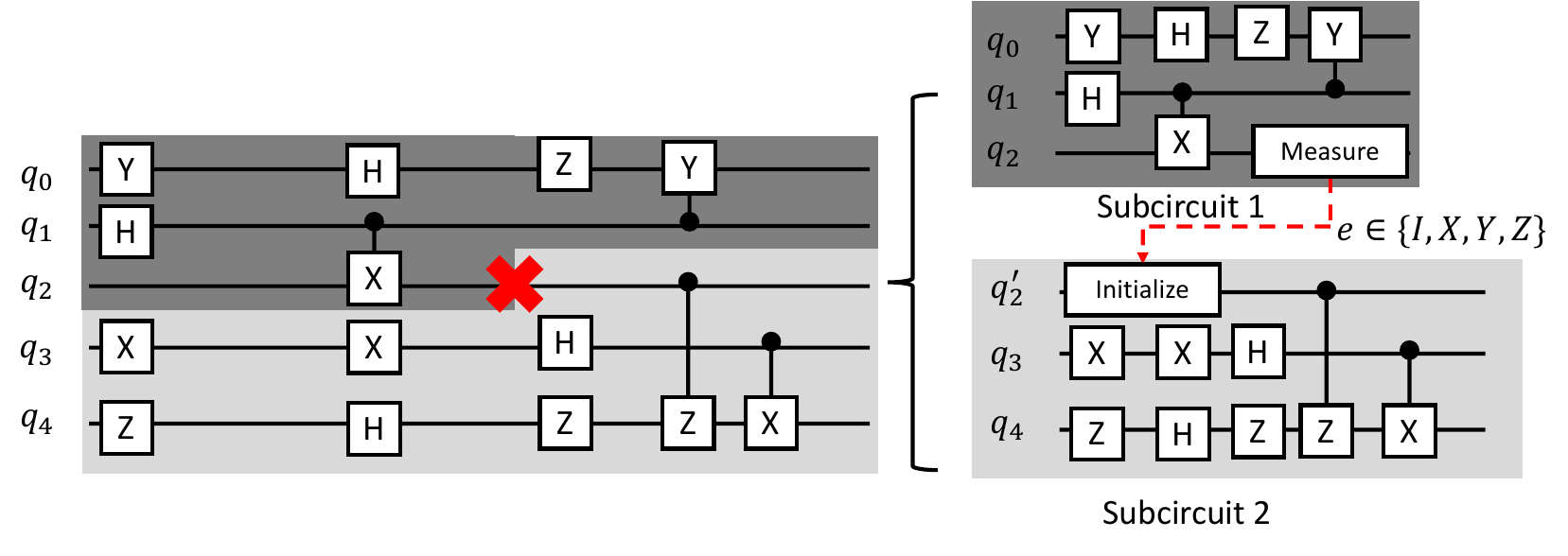}
    \caption{Example of cutting a $5$-qubit quantum circuit with one cut to divide it into two smaller subcircuits.
    (Left) The red cross indicates the cutting point.
    Subcircuit $1$ is shaded dark and subcircuit $2$ is shaded light.
    (Right) The dashed arrow between the subcircuits shows the path undertaken by the qubit wire being cut.
    The one cut needs to permute through the $\{I,X,Y,Z\}$ bases to reconstruct the unknown cut state.
    The two subcircuits require no quantum communications can now be executed independently in any order on multiple $3$-qubit QPUs.}
    \label{fig:cutting_example}
\end{figure*}

While we direct readers to~\cite{peng2020simulating} for a detailed derivation and proof of the physics theory,
we provide an intuitive understanding of the cutting process to identify the key challenges.

Figure~\ref{fig:cutting_example} shows an example of cutting a simple quantum circuit.
The left panel shows a $5$-qubit quantum circuit.
Each horizontal line is a qubit wire.
Boxes incident on a single qubit wire are single qubit quantum gates.
Boxes incident on two qubit wires are $2$-qubit quantum gates.
Without cutting, this circuit requires a QPU with at least five good enough qubits to execute all the quantum gates before too many errors accumulate.

Circuit cutting divides a large quantum circuit into smaller subcircuits.
For instance, a cut marked by a red cross can split the circuit into two distinct subcircuits.
This allows multiple less powerful $3$-qubit QPUs to run these independent subcircuits in parallel and in any sequence,
as there are no quantum connections necessary between them.
The process involves making vertical cuts along the qubit wires.
Generally, multiple cuts are used to separate a large quantum circuit into several smaller subcircuits.

Circuit cutting decomposes any unknown quantum states at cut points into a linear combination of their Pauli bases.
Specifically, QPUs run four variations of subcircuit $1$, each measures the upstream cut qubit $q_2$ in one of the $\{I,X,Y,Z\}$ Pauli bases.
One detail to note is that measuring in $I$ and $Z$ bases uses the same single-qubit rotations,
hence only $3$ subcircuits per upstream cut qubit are needed.
Correspondingly, QPUs run four variations of subcircuit $2$,
each initializes the downstream cut qubit $q_2'$ in one of the $\{\ket{0},\ket{1},\ket{+},\ket{i}\}$ states.
The $\{I,X,Y,Z\}$ Pauli bases are further constructed as a linear combination from the initialization states.
Measuring (initializing) a qubit in different bases simply means appending (prepending) various single qubit rotations on the qubit,
which are standard operations with little overhead and does not complicate the subcircuits.
Overall, the procedure produces four subcircuit $1$ outputs of $p_1^e$, for cut edge $e\in\{I,X,Y,Z\}$,
and similarly for $p_2^e$.

Circuit cutting involves classically reconstructing the quantum state outputs by combining the outputs of the subcircuits.
Prior works~\cite{peng2020simulating,tang2021cutqc} demonstrate that the binary state probability distribution $P$ of the original uncut circuit is equivalent to $\sum_{e}p_1^e\otimes p_2^e$.
This process effectively replaces quantum interactions between subcircuits with classical co-processing.

We introduce the following notations to describe the general quantum circuit cutting scenario:
\begin{enumerate}
    \item An $n$-qubit quantum circuit undergoes multiple cuts $e\in E$,
    for a total of $|E|$ cuts.
    \item Cuts in $E$ divide the input circuit into $m$ completely separated subcircuits $C_i\in\left\{C_1,\ldots,C_{m}\right\}$.
    \item $E_i\subseteq E$ represents the subset of cut edges in $E$ on $C_i$.
    In general, a subcircuit $C_i$ does not touch all the cut edges.
    Note that $C_i$ is only initialized and measured differently if $e\in E_i$ changes bases.
    \item $p_i^{\{e\}}$ represents the binary state probability output of $C_i$ for a particular set of cut edge bases $\forall e\in E_i$.
\end{enumerate}

The physics theory~\cite{peng2020simulating} dictates that the binary state probability vector output $P$ of the original $n$-qubit circuit is given by:
\begin{equation}
    P=\textcolor{blue}{\sum_{\{e_0\ldots e_{|E|-1}|e_i\in\{I,X,Y,Z\}\}}\otimes_{j=1}^{m}}\textcolor{red}{p_j^{\{e|e\in E_j\}}}\in\mathbb{R}^{2^n}\label{eq:reconstruction}
\end{equation}
where $\otimes$ is the tensor product between a pair of subcircuit binary state output vectors.
The red part of equation~\ref{eq:reconstruction} represents the QPU computations,
and the blue part of the equation represents the classical co-processing.
Equation~\ref{eq:reconstruction} shows that a full reconstruction permutes each cut edge $e$ through the $\{I,X,Y,Z\}$ bases,
for a total of $4^{|E|}$ permutation combinations.

\subsection{Circuit Cutting Challenges}
Equation~\ref{eq:reconstruction} clearly demonstrates the key challenges of applying quantum circuit cutting at useful scales:

\begin{enumerate}
    \item The classical co-processing overhead scales exponentially with the number of cuts as $4^{|E|}$ and hence bottlenecks the runtime.
    \item Reconstructing the full binary state probability distribution $P$ computes an $\mathbb{R}^{2^n}$ vector,
    which scales exponentially.
    \item Identifying high-quality cuts that reduce the classical co-processing overhead is crucial;
    however, this task constitutes an NP-complete constrained graph partition problem when applied to arbitrary quantum circuits.
\end{enumerate}

TensorQC combines frontier classical techniques in tensor networks and quantum computing to address the co-processing scalability challenges.
In addition, TensorQC models the hybrid runtime of distributing quantum circuits and proposes graph partition heuristics to automatically find cuts.
\section{Tensor Network Co-processing}\label{sec:co_processing}
This section explores the use of tensor network contraction for efficient classical co-processing.
Section~\ref{sec:prior_co_processing} discusses the limitations and inefficiencies of current parallel reconstruction methods.
Section~\ref{sec:tensor_network} introduces tensor networks and validates the mathematical correctness of using tensor network contractions for co-processing.
Section~\ref{sec:contraction_cost} evaluates the computational costs associated with tensor network contractions, demonstrating their exponential advantage over existing methods. 
Finally, Section~\ref{sec:tensor_network_compilation} details the compilation and execution processes for tensor networks.
\subsection{Why: Co-processing Inefficiencies in Prior Works}\label{sec:prior_co_processing}
\begin{figure}[t]
    \centering
    \includegraphics[width=\linewidth]{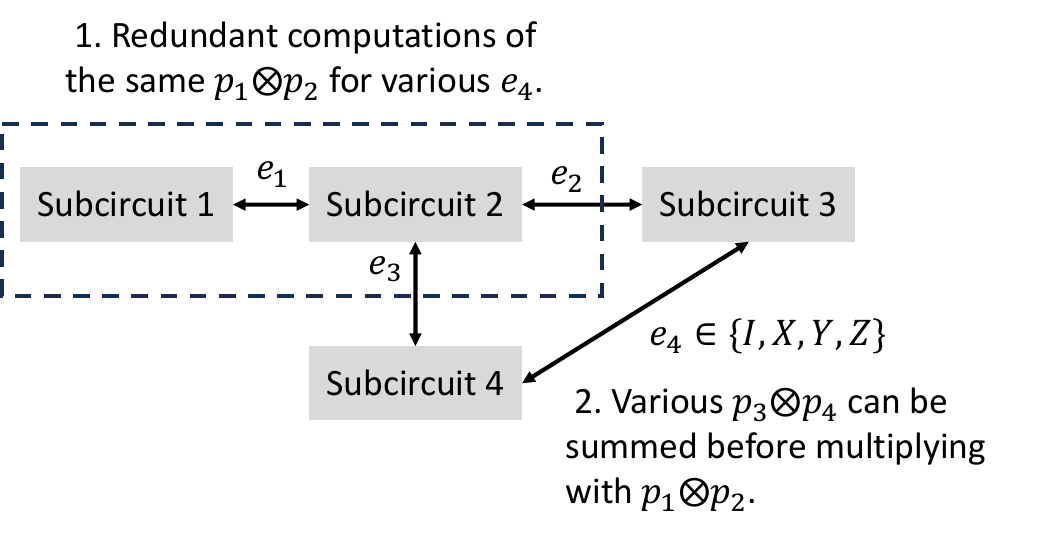}
    \caption{A hypothetical $4$-subcircuit scenario to demonstrate the two sources of inefficiencies of prior works.
    Consider the case where $e_{1,2,3}$ are fixed and only $e_4$ permutes.}
    \label{fig:prior_inefficiencies}
\end{figure}
State-of-the-art circuit cutting techniques~\cite{tang2021cutqc} follow the reconstruction formula in Equation~\ref{eq:reconstruction}.
This formula iteratively cycles through each of the ${I, X, Y, Z}$ bases for the $|E|$ cut edges,
resulting in a total of $4^{|E|}$ summation terms.
Each summation term requires computing the tensor products across the $m$ subcircuits.
The calculation of these summation terms is embarrassingly parallel,
and allows for straightforward distribution across CPUs and GPUs~\cite{tang2021cutqc,tang2022cutting}.
However, simple parallelization only reduces the runtime linearly in a fundamentally exponential problem.

The previous method incurs excessive costs due to redundant computations.
Typically, an input benchmark is divided into more than two subcircuits,
and not all subcircuits are interconnected by the cuts.
Since the output of a subcircuit $p_j^{\{e|e\in E_j\}}$ only alters when at least one of its connecting cuts changes bases,
many intermediate results remain unchanged and could be reused.
However, the earlier approach overlooks this possibility and redundantly recalculates all subcircuit products from scratch for each iteration.

Figure~\ref{fig:prior_inefficiencies} visualizes an example to demonstrate the two sources of inefficiencies of the prior method.
Specifically, the reconstruction formula~\ref{eq:reconstruction} translates to:
\begin{align}
    P&=\sum_{\{e_{1,2,3,4}\}}\otimes_{j=1}^{m}p_j^{\{e|e\in E_j\}}\nonumber\\
    &=\sum_{\{e_{1,2,3}\}}\sum_{\{e_4\}}p_1^{e_1}\otimes p_2^{e_{1,2,3}}\otimes p_3^{e_{2,4}}\otimes p_4^{e_{3,4}}\label{eq:prior_compute_example}
\end{align}
Consider the scenario described by Equation~\ref{eq:prior_compute_example} with fixed edge bases on $e_{1,2,3}$.
Firstly, the intermediate result $p_1^{e_1}\otimes p_2^{e_{1,2,3}}$ is utilized multiple times as the basis for $e_4$ changes.
The baseline method inefficiently recalculates the product from scratch for each $e_4$ permutation.
Secondly, the baseline method separately multiplies the same intermediate result, $p_1^{e_1}\otimes p_2^{e_{1,2,3}}$,
with four distinct $p_3^{e_{2,4}}\otimes p_4^{e_{3,4}}$ results corresponding to each possible state of $e_4$.
This method of repetitive and separate multiplications followed by summations,
is computationally inefficient and does not optimize the use of computing kernels.
A more efficient strategy involves using the distributive property of multiplication to restructure Equation~\ref{eq:prior_compute_example} as follows:
\begin{equation}
    \sum_{\{e_{1,2,3}\}}p_1^{e_1}\otimes p_2^{e_{1,2,3}}\otimes\sum_{\{e_4\}}p_3^{e_{2,4}}\otimes p_4^{e_{3,4}}
\end{equation}
By computing the summation of $p_3 \otimes p_4$ prior to its multiplication with $p_1 \otimes p_2$,
this method significantly reduces the number of multiplications required - from four to just one.
This strategic reordering simplifies the computational process by minimizing redundancy.
\subsection{What: Adapting Tensor Network Contraction}\label{sec:tensor_network}
\begin{figure}[t]
    \centering
    \begin{subfigure}{0.4\textwidth}
        \centering
        \includegraphics[width=.8\textwidth]{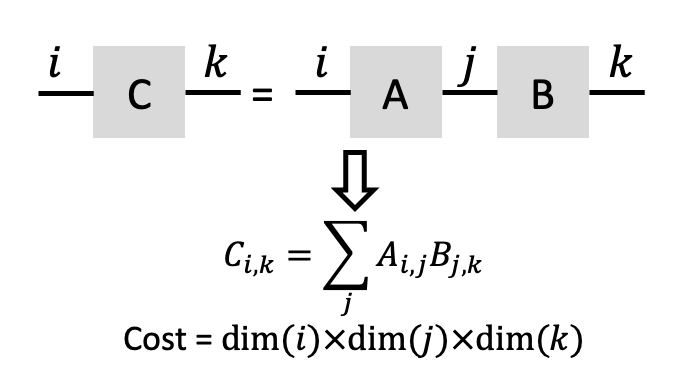}
        \caption{A pairwise tensor contraction with $1$ shared index $j$,
        which is the inner dimension being contracted.
        $i,k$ are the outer dimensions of the resulting big tensor $C$.
        The contraction cost is defined as the number of multiplications required.}
        \label{fig:tensor_contraction}
    \end{subfigure}
    \begin{subfigure}{0.4\textwidth}
        \centering
        \includegraphics[width=\textwidth]{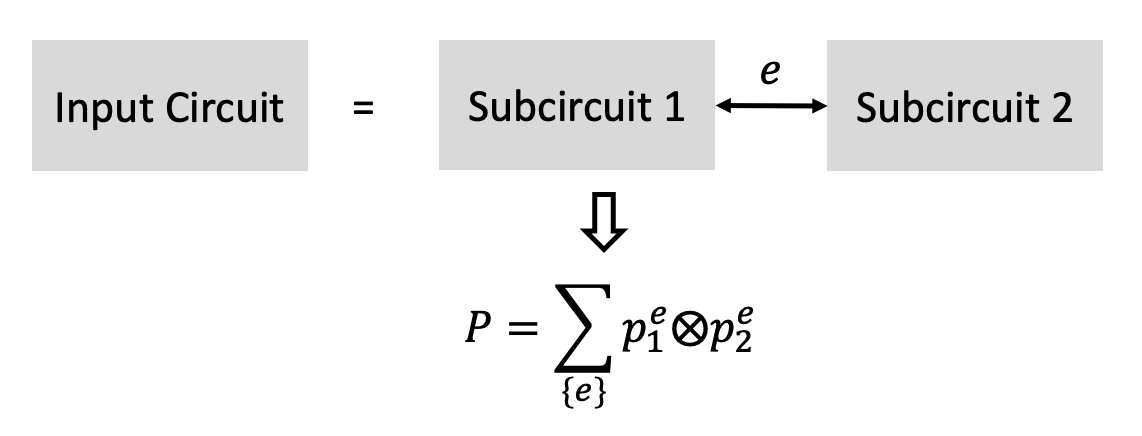}
        \caption{Reconstructing a pair of subcircuits means multiplying and summing over the cut edges in between.}
        \label{fig:subcircuit_contraction}
    \end{subfigure}
    \caption{Reconstructing two subcircuits is equivalent to a pairwise tensor contraction.}
    \label{fig:tensor_network_equivalence}
\end{figure}
Tensor networks have been widely used in classical simulations of quantum systems~\cite{vidal2003efficient,vidal2004efficient,schollwock2011density,verstraete2004matrix,tindall2023efficient}.
Tensors are essentially multidimensional arrays.
For example,
scalars are $0$-dimensional tensors,
vectors are $1$-dimensional tensors,
matrices are $2$-dimensional tensors, and so on.
Additionally, the indices of a tensor serve as coordinates that pinpoint the location of elements within the tensor.
The dimension of a tensor index indicates the number of elements along a specific axis.

A tensor network is a graphical representation that decomposes a large tensor into a product of smaller tensors interconnected through their indices.
Figure~\ref{fig:tensor_contraction} illustrates this concept with a tensor network that represents a large tensor $C$ using two smaller tensors $A$ and $B$.
In this example, $i, j, k$ denote the tensor indices.
Tensor $A$ has dimensions $\dim(i) \times \dim(j)$, tensor $B$ has dimensions $\dim(j) \times \dim(k)$,
and the resultant tensor $C$ has dimensions $\dim(i) \times \dim(k)$.
This specific network features a single shared index $j$ between $A$ and $B$.
Generally, pairs of tensors in a network can share multiple indices,
which facilitates more complex interactions and dimensional relationships within the network.

Tensor contraction involves merging smaller tensors into a larger tensor through multiplication and summation along their shared indices.
Figure~\ref{fig:tensor_contraction} provides a straightforward example of this process using just two tensors, $A$ and $B$.
Specifically, to contract tensors $A$ and $B$,
their elements are multiplied and then summed over the common index $j$.
The outcome of this contraction is a larger tensor $C$,
which is indexed by $i$ and $k$.
Furthermore, the computational cost of this operation,
in terms of floating point multiplications,
is determined by the product $\dim(i) \times \dim(j) \times \dim(k)$,
representing the combined dimensions of the indices involved in the contraction.

Figure~\ref{fig:subcircuit_contraction} illustrates the process of reconstructing two subcircuits,
treating each subcircuit output as a tensor indexed by the cut edges $e\in E_j$.
Varying the bases on these cut edges generates different output vectors $p_j^{\{e\}}$ for each subcircuit.
Notably, each cut edge is associated with a dimension of $4$,
corresponding to the permutation of bases $\{I,X,Y,Z\}$.
The reconstruction of a pair of subcircuits involves adjusting the bases on the shared cut edges,
multiplying the respective outputs,
and summing the results.
This process of combining the outputs from two subcircuits is mathematically analogous to contracting a pair of tensors,
where the tensors represent the outputs of the subcircuits.

This equivalence can be extended to any number of subcircuits.
Typically, a tensor network comprising $m$ tensors corresponds to dividing a circuit into $m$ subcircuits.
The shared indices between these tensors represent the cut edges connecting the subcircuits.
The process of multiplying the outputs from these subcircuits is mathematically analogous to contracting the tensor network,
which involves a sequence of $m-1$ pairwise contractions.
Similarly, the reconstruction of subcircuits entails sequentially combining pairs of subcircuits until all are merged into a single output.
\subsection{Tensor Network Contraction Cost}\label{sec:contraction_cost}
\begin{figure}[t]
    \centering
    \begin{subfigure}{0.4\textwidth}
        \centering
        \includegraphics[width=\textwidth]{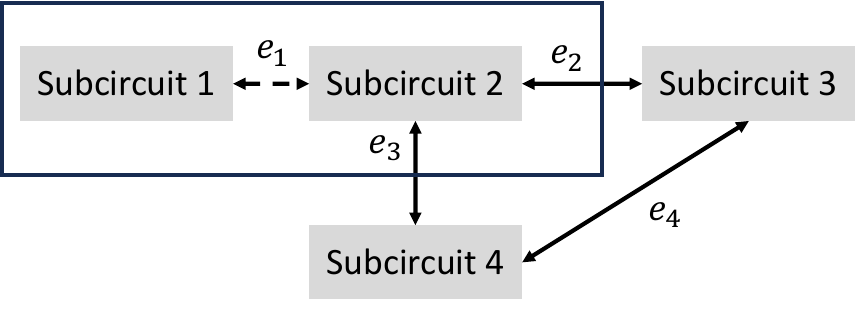}
        \caption{Contracting subcircuits $1$ and $2$ requires $4^1\times4^2=64$ multiplications.}
        \label{fig:contraction_step_1}
    \end{subfigure}
    \begin{subfigure}{0.4\textwidth}
        \centering
        \includegraphics[width=\textwidth]{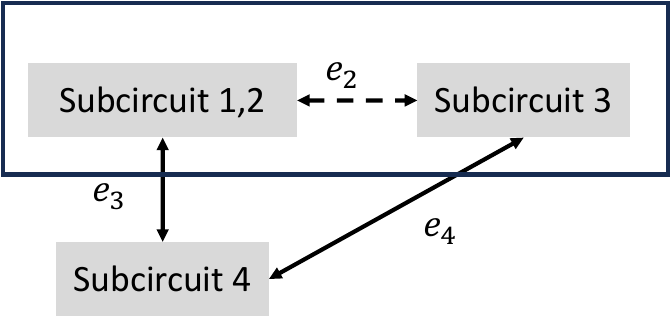}
        \caption{Contracting the already contracted subcircuits $1,2$ and $3$ requires $4^1\times4^1\times4^1=64$ multiplications.}
        \label{fig:contraction_step_2}
    \end{subfigure}
    \begin{subfigure}{0.2\textwidth}
        \centering
        \includegraphics[width=\textwidth]{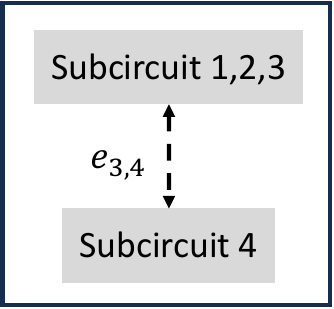}
        \caption{Contracting the already contracted subcircuits $1,2,3$ and $4$ requires $4^2=16$ multiplications.}
        \label{fig:contraction_step_3}
    \end{subfigure}
    \caption{Contracting the tensor network example from Figure~\ref{fig:prior_inefficiencies}.
    The solid boxes show the pair of subcircuits being contracted.
    The dashed edges inside the boxes are the inner dimensions at every contraction.
    The edges across the boundary of the boxes are the outer dimensions at every contraction.
    Tensor network contraction only requires $144$ multiplications in total.}
    \label{fig:contraction_steps}
\end{figure}

The computation cost of the co-processing is captured by the number of floating point multiplications.
Consider the computation cost to reconstruct one quantum state from the subcircuits in Figure~\ref{fig:prior_inefficiencies} using the prior method.
The prior method separately computes $4^4$ cut edge bases.
Each edge base permutation is a product of four subcircuits,
hence three multiplications.
Consequently, the prior method requires $4^4\times3=768$ floating point multiplications.

On the other hand,
Figure~\ref{fig:contraction_steps} demonstrates the application of tensor network contraction to process the same subcircuits.
This method involves sequentially multiplying subcircuit outputs until all are fully contracted.
According to Figure~\ref{fig:tensor_network_equivalence},
the cost of contracting a pair of subcircuits is determined by the product of the dimensions of all the cut edges.
With each cut edge cycling through four bases,
$\dim(e_i)=4,\forall i\in\{1,2,3,4\}$.
As a result, tensor network contraction of the same subcircuits requires only $144$ multiplications,
thereby reducing the computational expense by more than fivefold in this straightforward example.

Tensor networks offer a substantial reduction in the reconstruction costs associated with circuit cutting.
However, minimizing the overhead of tensor contraction remains challenging due to its complexity.
A general tensor network with $|E|$ edges can be contracted in $|E|!$ different sequences.
Although every sequence ultimately yields the correct result,
the associated contraction costs vary significantly.
Indeed, it is not uncommon for some sequences to be intractable,
while others are many orders of magnitude cheaper to compute.
The cost of contraction is influenced by several factors,
including the structure of the network and the dimensions of the tensors involved.
Regrettably, determining the optimal contraction sequence is an NP-hard problem,
making it computationally challenging to identify the most efficient approach~\cite{markov2008simulating, arnborg1987complexity, chi1997optimizing}.

Calculating an upper bound for the contraction complexity during subcircuit reconstruction is relatively straightforward.
Figure~\ref{fig:contraction_steps} shows that the total contraction cost is the sum of the costs of each individual contraction step.
The cost for each step is determined by the combined dimensions of both the internal and external cut edges of a tensor pair,
known as the \emph{contraction edges}.
Therefore, the cost for each contraction step is $4^{\#cuts}$,
where $\#cuts$ represents the total number of internal and external cuts involved in each contraction step.
Consequently, the overall contraction cost is upper-bounded by the following expression:
\begin{equation}
    \mathcal{O}(4^{K_{\max}}m)\label{eq:tn_upper_bound}
\end{equation}
where $K_{\max}$ is the max number of cut edges at any contraction step,
and $m$ is the number of subcircuits.
In contrast, the computation cost of the prior method is given by:
\begin{equation}
    \mathcal{O}(4^{|E|}m)\label{eq:prior_complexity}
\end{equation}

\subsection{Exponential Advantage from Tensor Network}
Let us analyze the exponential benefits of tensor network contractions through the following scenarios:
\begin{enumerate}
    \item Two subcircuits: $|E|=K_{max}$ since all cuts are on the two subcircuits and there is only one contraction step.
    The tensor network contraction cost is the same as ~\cite{tang2021cutqc}.
    \item Three subcircuits: The first tensor contraction step involves all cut edges and $K_1=|E|$.
    However, the second contraction step has $K_2<|E|$.
    Tensor contraction's total cost is hence $4^{K_1}+4^{K_2}$,
    while~\cite{tang2021cutqc} is $2\times4^{|E|}$.
    This results in cost savings, though not on an exponential scale.
    \item General cases with more than three subcircuits:
    Since an edge connects only two subcircuits,
    it follows that $K_1<|E|$ for the first contraction.
    For subsequent contractions, $K_i<|E|$ as some edges have already been contracted
\end{enumerate}
Therefore, $K_{\max}<|E|$ when there are more than three subcircuits.
Tensor network contraction hence introduces an exponential computation advantage over the prior method in general cases.
\subsection{How: Tensor Network Compilation}\label{sec:tensor_network_compilation}
\subsubsection{Determining Contraction Order}\label{sec:determining_contraction_order}
In the context of tensor network contraction,
the order in which subcircuits are contracted does not impact the accuracy,
yet it significantly influences computational overhead.
Although determining the optimal contraction sequence for a tensor network is NP-hard,
numerous heuristics have been developed to identify high-quality sequences,
making it a vibrant area of research across various fields~\cite{robertson1991graph,markov2008simulating,orus2014practical}.
For instance, TensorQC leverages the CoTenGra software~\cite{gray2021hyper} to optimize the contraction order,
enhancing performance without compromising output integrity.

\subsubsection{Managing Memory Constraints}
\begin{figure}[t]
    \centering
    \includegraphics[width=\linewidth]{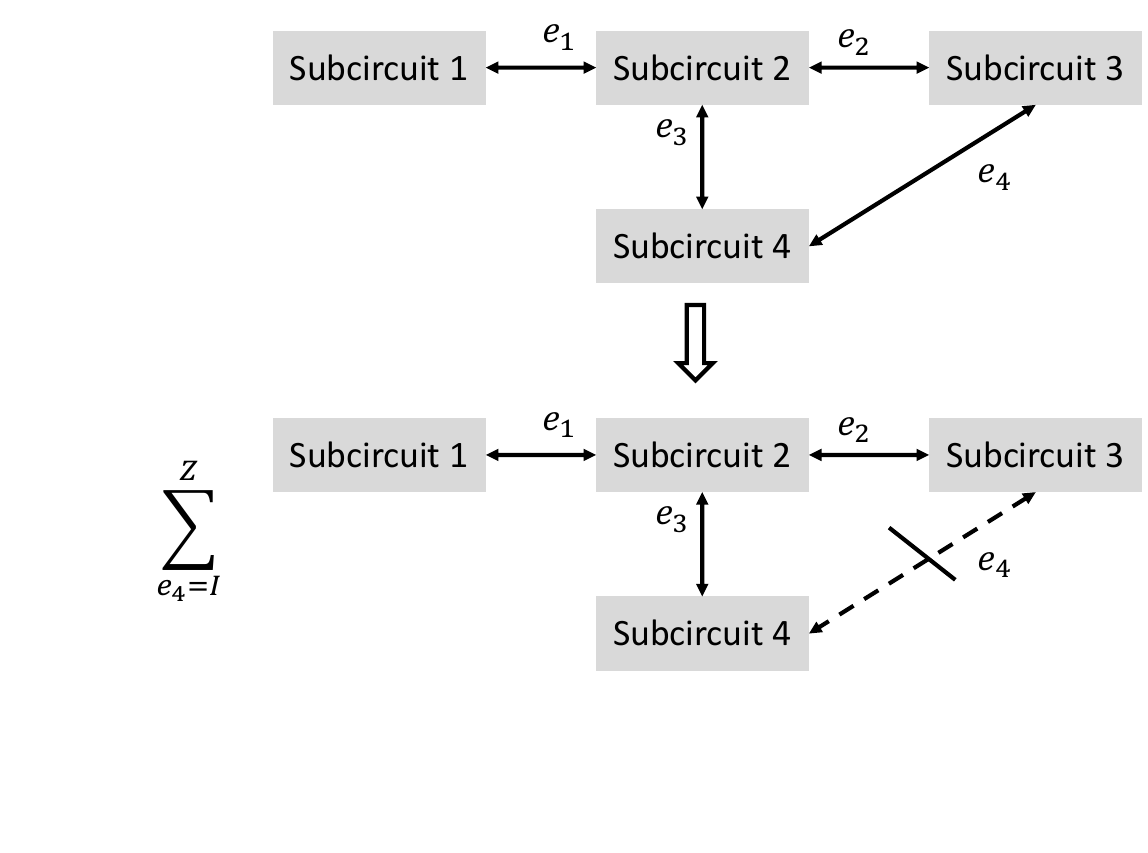}
    \caption{Slicing one cut edge to express a tensor network as a summation of smaller tensor networks.}
    \label{fig:slicing}
\end{figure}

To ensure that the computations are both efficient and feasible within the hardware limitations,
it is crucial that the input subcircuit tensors fit within the GPU's memory constraints.
TensorQC implements an index slicing strategy that breaks down some cut edges to decompose the full tensor network into a summation of smaller tensor networks,
thereby reducing the size of the tensors to be processed.
For example, slicing the edge $e_4$ (as shown in Figure~\ref{fig:slicing}) results in four smaller tensor networks,
each representing a different fixed basis of $e_4$.
This method allows TensorQC to manage memory usage effectively by continuously slicing the most size-reducing cut edge until the tensor sizes are compatible with the GPU's memory capacity.

Moreover, to prevent memory overflow during the tensor contraction process,
a second-level slicing strategy is employed.
This method further divides the indices of individual tensor networks that were sliced in the previous step,
ensuring that both input and intermediate tensors remain within memory limits.
This two-tiered approach to tensor slicing ensures that all components fit within the available GPU memory.
\section{Heavy State Selection}\label{sec:heavy_state_selection}
\begin{algorithm}[t]
    \DontPrintSemicolon
    \SetAlgoLined
    \caption{Heavy State Selection}\label{alg:heavy_state_selection}
    \KwIn{Subcircuit outputs $p_{j,i}^{\{e|e\in E_j\}}$ for subcircuit $C_j$, edge permutation $e$ and binary state $i$.
    Maximum number of states $s_{\max}$.}
    Initialize $x_j=\emptyset$, $\forall j$.\;
    \ForEach{$C_j$, $i$}{
        Compute the $L2$ norm for all edge permutations $||p_{j,i}||\equiv\sqrt{\sum_{e}(p_{j,i}^e)^2}$.\;
        Add $arg\max||p_{j,i}||$ to $x_j$.\;
        Remove $\max||p_{j,i}||$.\;
    }
    Sort $||p_{j,i}||$ in descending order.\;
    \While{$\prod_j|x_j|<s_{\max}$}{
        Pop $||p_{j,i}||$ and select binary state $i$ of $C_j$.\;
        Add $i$ to $x_j$.\;
    }
    \Return{Heavy states $x_j$ for each $C_j$}
\end{algorithm}

While tensor network contraction significantly reduces the co-processing overhead,
it still incurs significant memory overhead to reconstruct the full $2^n$ Hilbert space.
When running QC applications, users do not seek every single binary state amplitude,
but rather are only interested to know the ones with higher amplitudes.
TensorQC proposes an automatic Heavy State Selection (HSS) protocol to only reconstruct the most important binary states.
This technique greatly reduces the co-processing overhead,
while retaining  most of the amplitude weights.

Algorithm~\ref{alg:heavy_state_selection} outlines the HSS algorithm to prune subcircuit outputs prior to tensor network contraction.
For a given subcircuit entry $p_j^{\{e\in E_j\}}$,
HSS computes the $L2$ norm of every binary state,
and retains the subcircuit binary states with higher weights.
HSS hence only retains the most significant subcircuit output binary states.
Suppose HSS ends up retaining $|x_j|$ binary states for each subcircuit $C_j$,
the output dimension reduces from $2^n$ to $\prod_j |x_j|$.
\section{Hybrid Cost Model}\label{sec:cost_model}
DHQC incorporates both quantum and classical computational costs,
presenting a trade-off between the complexity of tensor networks and the quantum resources required.
Numerous cuts lead to complex tensor networks,
increasing classical costs but also producing smaller subcircuits that require smaller QPUs.
Conversely, fewer cuts may not significantly reduce subcircuit sizes,
but result in simpler tensor networks that are easier to contract.

To estimate QPU runtime,
we assume the following operational times using IBM QPU data~\cite{ibm_quantum}:
\begin{enumerate}
    \item One layer of quantum gates runs in $t_g=10^{-7}$ seconds.
    \item Quantum measurements take $t_m=10^{-6}$ seconds.
\end{enumerate}

Additionally, TensorQC configures the number of shots equal to the number of quantum states per subcircuit,
with limits ranging from $2^{10}$ to $2^{20}$.
We define the following parameters for subcircuits:
\begin{enumerate}
    \item $u_i$: Number of upstream cut qubits in subcircuit $C_i$.
    \item $d_i$: Number of downstream cut qubits in subcircuit $C_i$.
    \item $w_i$: Width or the number of qubits in subcircuit $C_i$.
    \item $t_i$: Depth or the number of quantum gate layers in subcircuit $C_i$.
\end{enumerate}
The total QPU runtime in seconds is calculated as:
\begin{equation}
    T_{QPU}\equiv \sum_{i=1}^m 3^{u_i}\times4^{d_i}\times2^{w_i}\times(t_i\times t_g + t_m)\label{eq:qpu_runtime}
\end{equation}
The subcircuits can be run in parallel since there is no quantum data dependency between them,
allowing for execution on multiple QPUs simultaneously.
Consequently, having more QPUs linearly decreases the QPU runtime.

The tensor network contraction sequence is optimized using heuristic algorithms to minimize classical computation costs,
quantified by the number of floating point operations (FLOPs).
The classical runtime is then estimated by dividing the total contraction cost by the computational throughput of the backend classical system:
\begin{equation}
    T_{classical}\equiv \frac{C_{TN}}{FLOPs}\label{eq:classical_runtime}
\end{equation}
where $C_{TN}$ is the cost of contracting a tensor network and $FLOPs$ is the floating point operations per second of a backend GPU.

Overall, TensorQC seeks to minimize the sum of QPU and classical co-processing runtimes.
\section{Locating Cutting Points}\label{sec:locate_cuts}
This section defines the task of identifying cuts within a quantum circuit as a constrained multi-way graph partitioning problem
and introduces a solution that employs a greedy graph growing algorithm.

\subsection{Problem Formulation}
\begin{figure}[t]
    \centering
    \begin{subfigure}{0.5\textwidth}
        \centering
        \includegraphics[width=\textwidth]{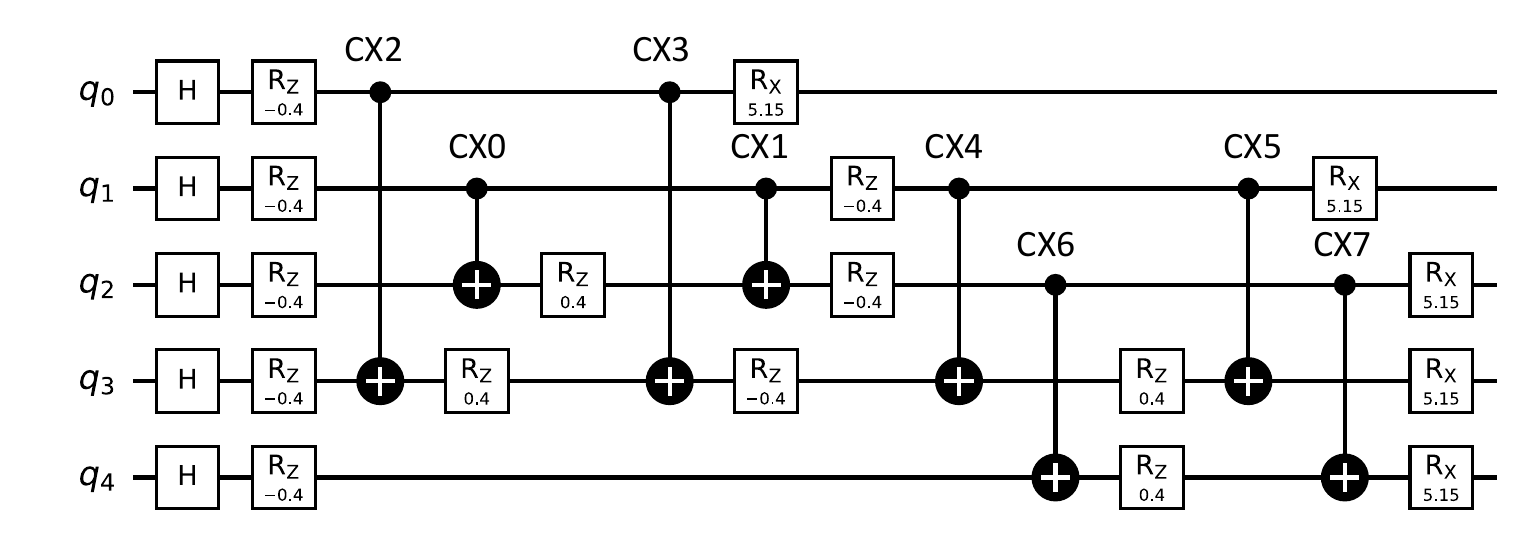}
        \caption{QAOA benchmark circuit solving the maximum independent set problem for a random Erdos-Renyi graph with $5$ qubits.}
        \label{fig:circuit}
    \end{subfigure}
    \begin{subfigure}{0.5\textwidth}
        \centering
        \includegraphics[width=\textwidth]{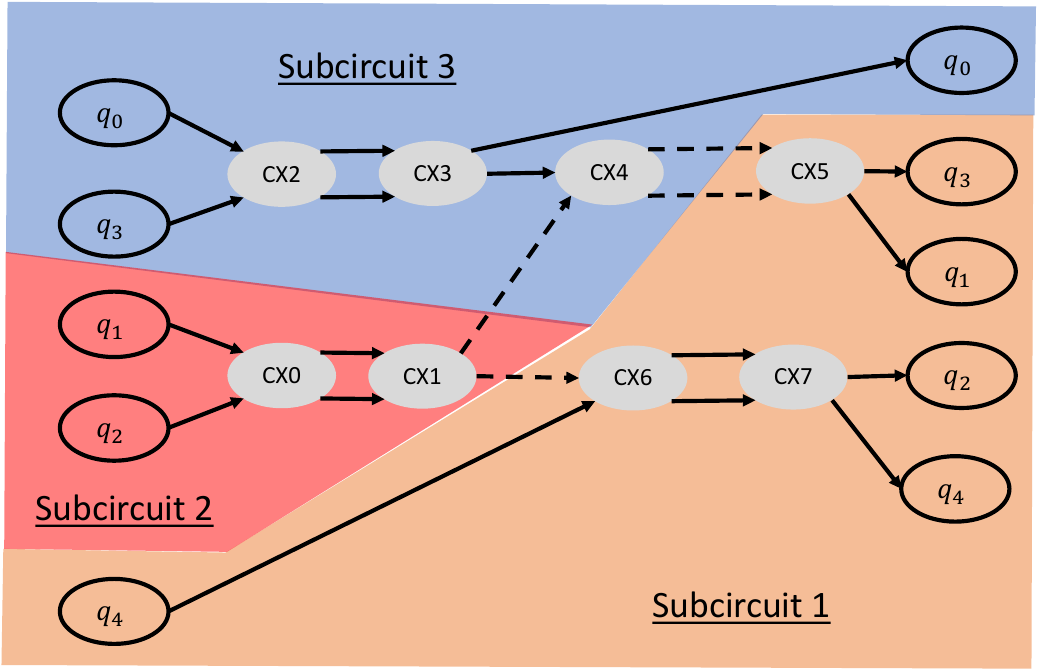}
        \caption{DAG representation of~\ref{fig:circuit} without the single-qubit gates.
        The shaded vertices indicate the two-qubit quantum gates.
        The directed edges indicate the time flow of the qubit wires.
        Qubits in the initial states (left column) evolve through the quantum gates and are measured as output (right column).
        The colored shades delineate an example partition.}
        \label{fig:cut_dag}
    \end{subfigure}
       \caption{Finding cuts for a quantum circuit reduces to finding a partition of the quantum gates.
       The dashed qubit wires across different partitions are the cut edges.}
       \label{fig:tensor_graph_example}
\end{figure}

\subsubsection{Reducing to Graph Partition}
Quantum circuits can be depicted as directed acyclic graphs (DAGs).
Figure~\ref{fig:circuit} illustrates a $5$-qubit quantum circuit,
while Figure~\ref{fig:cut_dag} displays its corresponding DAG representation.
In the DAG, single-qubit gates are omitted to clearly highlight the circuit's topology.
The vertices in the DAG represent the input/output qubits and gates,
and the edges denote the connections between these qubits and gates through qubit wires.

Identifying cuts within this framework is akin to partitioning the two-qubit quantum gates.
The placement of cut edges precisely determines the boundaries of subcircuits.
In the context of DAG partitioning,
the focus is primarily on how vertices are interconnected;
therefore, single-qubit gates,
which do not influence the overall topology,
are excluded from the cut finding process.
These gates are instead assigned to the same subcircuit as their adjacent two-qubit gate.
Additionally, input and output qubit vertices should not be isolated in the partitioning process.

\subsubsection{Constraints}
QPU hardware is constrained by both the limited size and quality of qubits,
presenting two constraints.
The available sizes of QPUs set strict upper limits on the width of each subcircuit.
Furthermore, as Noisy Intermediate-Scale Quantum (NISQ) QPUs are error-prone,
they can support only a limited number of quantum gates before too many errors accumulate in each subcircuit.
Looking ahead, while fault-tolerant QPUs are expected to ease the limitations on subcircuit size significantly,
they will still be subject to circuit width constraints.

\subsubsection{Problem Statement}
TensorQC aims to identify cuts that optimize contraction costs while adhering to the stringent limitations imposed by QPU hardware.
To this end, we formalize the problem of finding cuts within a quantum circuit as follows:

Consider the DAG $g$ representing a quantum circuit,
the maximum number of qubits $w_{\max}$ that an available QPU can handle,
and the maximum number of gates $s_{\max}$ that the QPU can support.
Assuming that partitioning the quantum gates in $g$ results in $m$ distinct partitions,
the task is to find a partition that meets the following criteria:
\begin{enumerate}
    \item $0<w_i\leq w_{\max}, \forall i\in \{1,\ldots,n\}$, where $w_i$ is the number of qubits in each partition.
    \item $0<s_i\leq s_{\max}, \forall i\in \{1,\ldots,n\}$, where $s_i$ is the number of gates in each partition.
\end{enumerate}
The objective is to minimize the total runtime:
\begin{equation}
    T\equiv T_{QPU} + T_{classical}\label{eq:cut_objective}
\end{equation}
\subsection{Greedy Graph Growing}

\begin{algorithm}[t]
    \DontPrintSemicolon
    \SetAlgoLined
    \caption{Greedy Graph Growing}\label{alg:graph_growing}
    \KwIn{Quantum circuit DAG $g$.
    Max number of qubits on QPUs $w_{\max}$.
    Max number of gates supported by QPUs $s_{\max}$.
    Threshold number of contraction edges $K_{t}$.
    Merging cost cutoff $Q_{\max}$.}
    Initialize a DAG $g'=[E,V]$ with only the $2$-qubit gates in $g$.\;
    Compute the cost of merging $\forall e\in E$ using Algorithm~\ref{alg:compute_merging_cost}.\;
    Select the edge $e_{\min}$ with the minimum merging cost for which $Q_e\leq Q_{\max}$.\;
    \While{$\exists e_{\min}$}{
        Merge the edge $e_{\min}$ to create a new partition.\;
        Update the merging costs of all neighboring edges on the new partition using Algorithm~\ref{alg:compute_merging_cost}.\;
        Select $e_{\min}$ if there are valid edges remaining.\;
    }
\end{algorithm}

\begin{algorithm}[t]
    \DontPrintSemicolon
    \SetAlgoLined
    \caption{Compute Edge Merging Cost}\label{alg:compute_merging_cost}
    \KwIn{DAG $g=[E,V]$.
    Subset $E'\subseteq E$.
    Various threshold values.}
    \For{$e\in E'$}{
        Merge the two vertices connected by $e$ in $g$.\;
        Count the number of qubits $w_{trial}$ and gates $s_{trial}$ in the newly merged trial partition.\;
        Count the max number of contraction edges $K_{trial}$ between the trial partition and its neighbors.\;
        Compute the sum of costs $Q_e$ from $w_{trial}$, $s_{trial}$ and $K_{trial}$ according to Table~\ref{table:merging_cost}.\;
        Un-merge $e$.\;
        }
\end{algorithm}

\begin{table}[t]
    \centering
    \begin{tabular}{ |c|c|c| } 
        \hline
         & Subcircuit Widths \& Sizes & \#Contraction Edges \\
        \hline
        $\leq T$ & $x/T$ & $x/T$ \\ 
        \hline
        $>T$ & Not allowed & $4^{x-T}+1$ \\ 
        \hline
       \end{tabular}
    \caption{The cost of merging an edge depends on the metric value $x$,
    its respective threshold $T$,
    and whether it is a quantum hard constraint or a classical penalty.
    Breaking quantum constraints thresholds is not allowed while breaking classical cost thresholds imposes heavy penalties.}
    \label{table:merging_cost}
\end{table}

Constrained graph partition problems,
which involve partitioning nodes in a way that satisfies certain conditions,
are believed to have no polynomial time solutions~\cite{andreev2006balanced, karypis1998fast}.
Prior works~\cite{tang2021cutqc} employed Mixed Integer Programming (MIP) to minimize the number of cuts in a quantum circuit.
However, this approach is notably slow and fails to efficiently scale beyond circuits with approximately $100$ qubits.
Furthermore, traditional graph partition heuristics like METIS~\cite{karypis1998multilevelk} only deal with static graphs.
These algorithms are not suitable for quantum circuits because the subcircuit qubit counts increase with each cut,
making them difficult to treat as static input variables.

Instead, TensorQC has developed heuristic methods to approximate the total runtime objective~\ref{eq:cut_objective}.
According to the tensor network contraction complexity formula~\ref{eq:tn_upper_bound},
the maximum number of contraction edges at any step in an optimal contraction sequence provides an upper bound on the cost.
Yet, identifying an optimal contraction sequence remains NP-hard,
making it challenging to determine the precise $K_{\max}$.

Exhaustive counting does not scale.
Instead, TensorQC simplifies the approach by counting the number of cut edges needed to contract any pair of neighboring subcircuits,
using this as a proxy for $T_{classical}$.
For example, Figure~\ref{fig:cut_dag} requires counting the number of cut edges if we were to contract
subcircuits $1,2$, subcircuits $1,3$, or subcircuits $2,3$.
This method scales effectively,
as it limits the counting to no more than $|E|$ pairs of neighboring subcircuits,
providing a reasonable estimate of contraction costs across various sequences.

In addition, the number of cuts on each subcircuit has a significant exponential impact on the quantum runtime $T_{QPU}$.
Hence, counting the cut edges also indirectly approximates the quantum runtime $T_{QPU}$.

To optimize the total runtime,
we introduce a threshold $K_{t}$ for the number of cut edges.
The heuristics heavily penalize any pair of neighboring subcircuits exceeding this threshold,
with $K_{i,j}>K_{t}, \forall i,j\in \{1,\ldots,m\}$,
where $K_{i,j}$ is the number of contraction edges for a pair of neighboring subcircuits.

Algorithm~\ref{alg:graph_growing} details the heuristic approach,
with Algorithm~\ref{alg:compute_merging_cost} serving as a subroutine.
At a high level, this algorithm evaluates the cost of merging potential edges and selects the most advantageous merger.
The merging process continues until further merging would exceed QPU capabilities or incur prohibitive classical costs.
The cost of a potential merge is calculated by aggregating both quantum and classical impacts,
as summarized in Table~\ref{table:merging_cost}.
\section{Methodology}\label{sec:methodology}

\subsection{Backends}\label{sec:backends}
The classical post-processing in the experiments runs on a single Nvidia $A10G$ GPU.
QPUs nowadays are still too small and noisy to even support running medium size subcircuits for meaningful analysis.
Instead, we use random numbers as the subcircuit output to focus on demonstrating the post-processing runtimes for large benchmarks.
In addition, we use noiseless classical simulators to study the HSS approximation effects for small benchmarks.
We expect more reliable QPUs to enable a full evaluation of TensorQC for practical applications.
We further assume distributing subcircuits to $10$ small QPUs to quantify the QPU runtime,
a reasonable assumption as IBM cloud provides around $15$ QPUs nowadays.
In practice, we set $K_t=10$ and $Q_{\max}=10^4$ for the Greedy Graph Growing heuristics.
\subsection{Benchmarks}\label{sec:benchmarks}
We used the following circuits as benchmarks:
\begin{enumerate}
    \item $Regular$: Quantum Approximate Optimization Algorithm (QAOA) solves the maximum independent set problem for random $3$-regular graphs~\cite{saleem2020approaches}.
    The full QAOA training process to optimize the hyperparameters is beyond the scope of this paper.
    We use random hyperparameters with the same circuit structure.
    Distributed landscape.
    \item $Erdos$: The same algorithm as $Regular$ but for random Erdos-Renyi graphs.
    Distributed landscape.
    \item $AQFT$: Approximate Quantum Fourier Transform~\cite{barenco1996approximate} that is expected to outperform the standard QFT circuit under noise,
    which is an essential subroutine in Shor's algorithm.
    Uniform landscape.
    \item $Supremacy$: Random quantum circuits adapted from~\cite{boixo2018characterizing}.
    Deeper versions of the circuit was used by Google to demonstrate quantum advantage~\cite{arute2019quantum}.
    Our paper uses depth $(1+8+1)$ at various sizes.
    Distributed landscape.
    \item $W\_State$: Generates the generalized W state~\cite{dur2000three, diker2016deterministic}.
    $n$ dominating states for a $n$-qubit benchmark.
    \item $GHZ$: Generates the entangled GHZ state~\cite{greenberger1989going}.
    Two dominating states.
\end{enumerate}
Our benchmarks cover a comprehensive range of circuit structures and quantum state entanglements.
TensorQC demonstrates running small benchmarks to analyze the fidelity,
and large benchmarks that are beyond the reach of any existing computation methods.
\subsection{Metrics}
There are three key metrics this paper looks at, namely runtime, quantum resources and accuracy.

\paragraph{Runtime, faster is better}
The runtime for purely QPU executions without cutting is simply the QPU time to run a circuit as is,
which is the fastest.
The TensorQC runtime is the combined quantum and classical runtime defined in~\ref{eq:cut_objective}.
Our experiments capture the real GPU wall clock runtime to reconstruct $1$ million arbitrary quantum states.

\paragraph{Quantum Area, smaller is better}
The quantum resource requirement is loosely defined as the product of the circuit width and depth,
called the `quantum area'.
Classical simulations require $0$ quantum area as it does not use QPUs at all.
For purely QPU executions, it is simply the product of the number of qubits and the circuit depth of the input quantum circuit.
For TensorQC, it is defined on the largest subcircuit produced from cutting.
The rationale is that QPUs must be able to support the workloads with a certain quantum area at high accuracy to produce accurate results.
While many hardware and software factors affect the QPUs' ability to support quantum workloads,
a smaller quantum area generally puts less burden on the quantum resources.

\paragraph{Amplitude Reconstruction Ratio, higher is better}
HSS only retains a subset of the output states,
hence does not account for the full probability amplitudes.
We represent the amount of amplitudes reconstructed as:
\begin{equation}
    P_{HSS}=\sum_{i\in \{HSS\}}P_i
\end{equation}
where $P_i$ is the reconstructed output of the full circuit,
$\{HSS\}$ indicates the set of binary states retained by HSS.
$P_{HSS}=1$ means HSS captures all the non-zero binary states.

In addition, different benchmarks have different max possible $P_{HSS}$
for a given number of sampled states,
noted by $P_{HSS}^{max}$.
This is captured by the sum of amplitudes of the top $|HSS|$ states in its error-free output:
\begin{equation}
    P_{HSS}^{max}\equiv\sum_{i=1}^{|HSS|}P_i^{\textrm{true, sorted}}
\end{equation}
where $P^{\textrm{true, sorted}}$ is the sorted error-free amplitude distribution of a benchmark circuit.
$P_{HSS}^{max}$ solely depends on the benchmark circuit itself and dictates the best possible performance of any sampling algorithms.
For example, $GHZ$ has a highly skewed output landscape with $2$ solution states sharing all the amplitudes.
Hence $P_{HSS}^{max}=1$ for $|HSS|\geq2$.
On the other hand, $Regular$, $Erdos$, $AQFT$, and $Supremacy$ benchmarks have more evenly distributed output landscapes,
without any dominating states.
Therefore, the ratio between the two represents the fraction of the max possible amplitude reconstructed given a certain $|HSS|$ limit,
ranging from $0$ to $1$:
\begin{equation}
    r_{P,HSS}\equiv P_{HSS}/P_{HSS}^{max}\label{eq:hss_amplitude_ratio}
\end{equation}

\paragraph{HSS Data Efficiency, higher is better}
We aim to quantify the data efficiency of the HSS protocol by stating,
“The HSS protocol retains $X\%$ of the total amplitudes while reconstructing only $Y\%$ of the binary states”,
where higher $X$ and lower $Y$ are preferred.

While $P_{HSS}$ represents $X$,
the number of reconstruction states $|HSS|$ as a fraction of the size of the Hilbert space for a $n$ qubit benchmark quantifies $Y$:
\begin{equation}
    r_{|HSS|}\equiv\frac{|HSS|}{2^n}
\end{equation}
Both $P_{HSS}$ and $r_{|HSS|}$ range between $0$ and $1$.
We therefore quantify the HSS data efficiency by the ratio between the two:
\begin{equation}
    \eta_{HSS}\equiv\frac{P_{HSS}}{r_{|HSS|}}\label{eq:hss_efficiency}
\end{equation}
Note that if $|HSS|=2^n$,
the HSS protocol reconstructs all the binary states and $P_{HSS}=r_{|HSS|}=1$.

Our goal for the HSS protocol is to achieve more amplitude reconstruction using fewer states,
a data efficiency represented by a higher $\eta_{HSS}$.
$\eta_{HSS}=1$ represents a one-to-one trade-off between the amplitude reconstruction and the number of states sampled.
\section{Experiment Results}\label{sec:results}
\subsection{Runtime}\label{sec:runtime_results}
\begin{figure}[t]
    \centering
    \includegraphics[width=\linewidth]{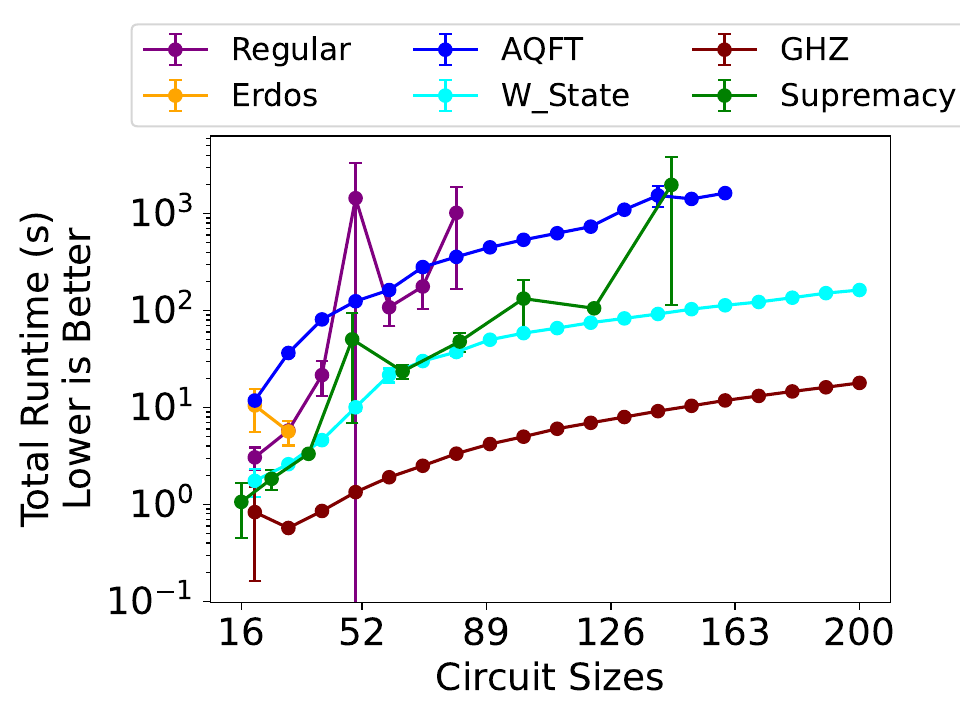}
    \caption{End-to-end wall clock runtimes.
    Each data point is the average of $3$ trials.
    The error bars represent the standard deviations.
    Experiments terminate benchmarks when the estimated tensor contraction runtime exceeds $10^3$ seconds.}
    \label{fig:total_runtime}
\end{figure}

\begin{figure}[t]
    \centering
    \includegraphics[width=\linewidth]{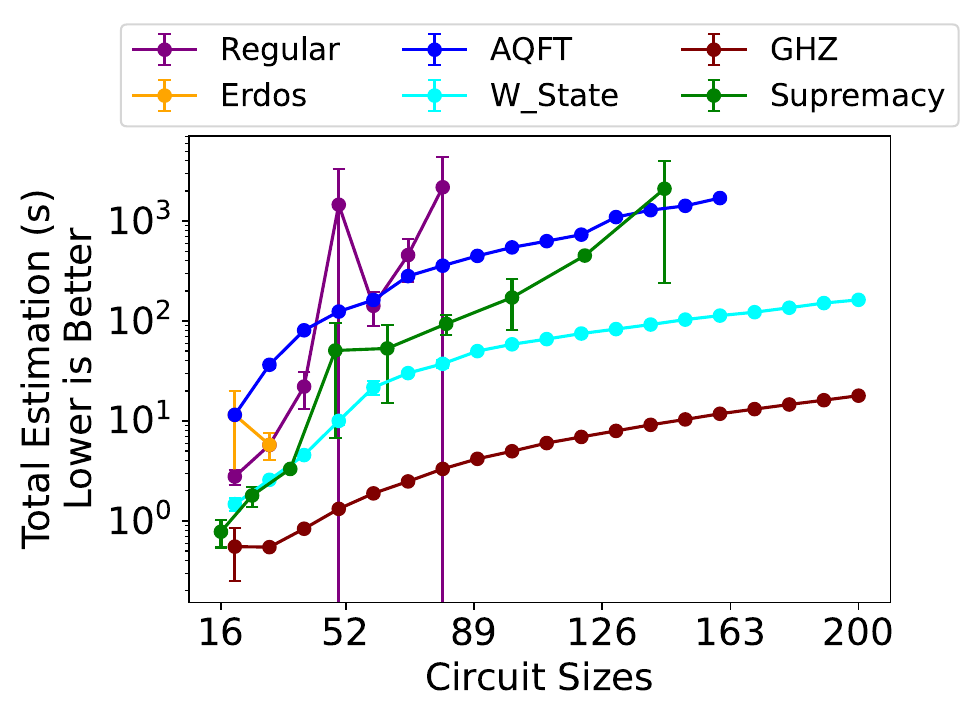}
    \caption{End-to-end runtime estimations using Equation~\ref{eq:cut_objective} for the same experiments in Figure~\ref{fig:total_runtime}.
    The estimated runtimes closely match with experimental data.}
    \label{fig:total_runtime_estimation}
\end{figure}

\begin{figure}[t]
    \centering
    \includegraphics[width=\linewidth]{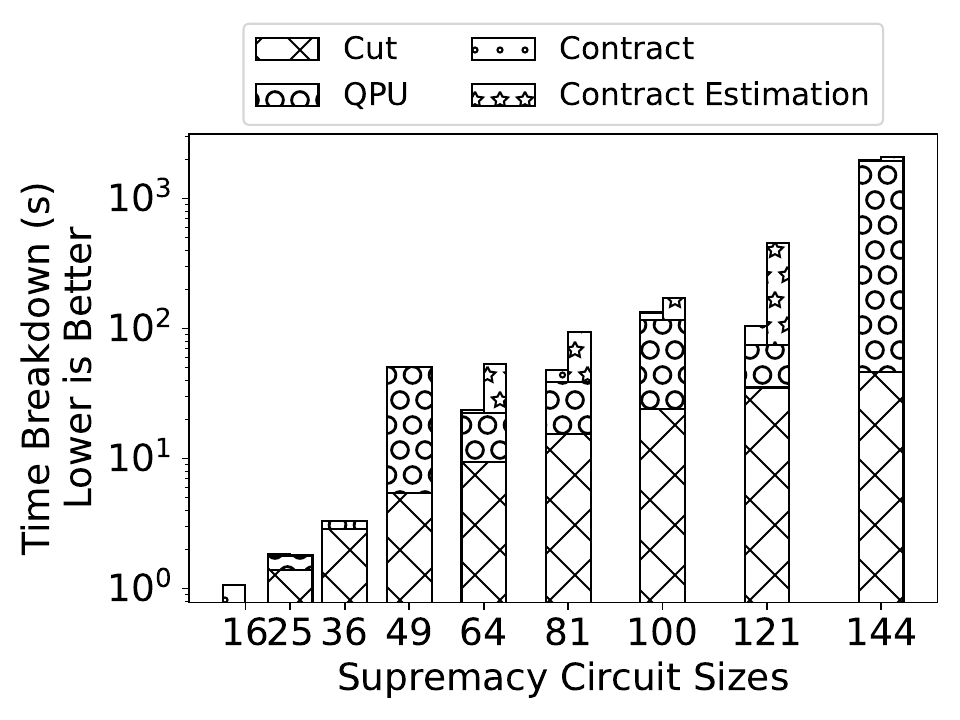}
    \caption{The runtime breakdown for the $Supremacy$ benchmarks.}
    \label{fig:runtime_breakdown}
\end{figure}

Figure~\ref{fig:total_runtime} shows the end-to-end wall clock runtime of computing $1$ million states for various benchmarks.
Experiments limit each benchmark circuit to at most half the number of qubits and gates in the uncut benchmark.
Note that these subcircuit size limits are upper bounds used in searching for cuts,
rather than the max subcircuit size produced from cutting.
Figure~\ref{fig:quantum_area} shows the actual quantum area found from cutting for the same set of experiments.

Figure~\ref{fig:total_runtime_estimation} shows the end-to-end runtime estimations obtained from Equation~\ref{eq:cut_objective}.
The estimations predict the tensor network contraction times using the heuristics number of floating point operations obtained from tensor network compilation,
divided by the backend GPU FLOPs.
The cut searching and QPU runtime estimations are the same as Figure~\ref{fig:total_runtime}.
Hence, estimations using Equation~\ref{eq:cut_objective} tend to overestimate the actual wall clock runtime,
but accurately reflect the trends.

Figure~\ref{fig:runtime_breakdown} shows the wall clock and estimation runtime breakdown of the $Supremacy$ benchmarks.
The inaccuracy in the runtime predictions is from the estimated time to contract subcircuit tensor networks.
Equation~\ref{eq:classical_runtime} consistently overestimates the GPU runtime.
We conjecture that more accurate estimations further require a holistic profiling of tensor network contraction on specific backend GPUs.
In addition, our cut searching algorithm~\ref{alg:graph_growing} finds cut solutions that balance between the quantum and classical runtimes,
showcasing its ability to find sweet tradeoff points.
Despite the efficiency of the heuristics cut search algorithm,
cuts finding still takes a significant portion of the end-to-end runtimes.
\subsection{Exponential Classical Overhead Advantage}
\begin{figure}[t]
    \centering
    \includegraphics[width=\linewidth]{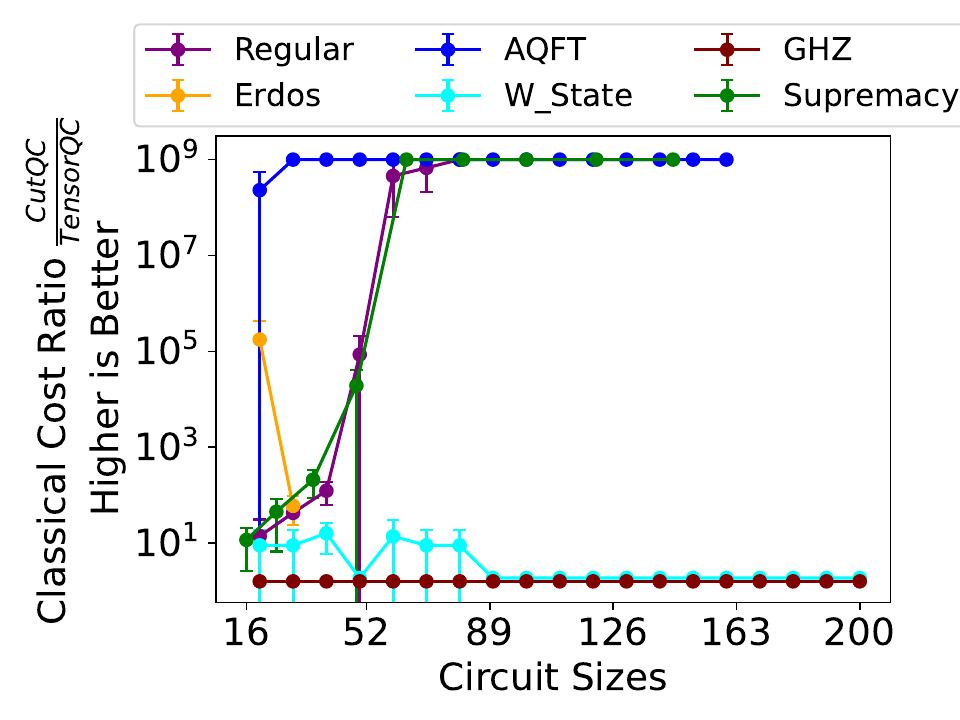}
    \caption{Classical post-processing overhead comparison of CutQC over TensorQC
    for the same experiments in Figure~\ref{fig:total_runtime} on a $\log$ scale.
    Max ratios are clipped at $10^9$ for clarity.}
    \label{fig:cost_ratio}
\end{figure}

Figure~\ref{fig:cost_ratio} shows the classical co-processing overhead of CutQC~\cite{tang2021cutqc} over TensorQC.
Specifically, we compute the number of scalar multiplications for TensorQC's tensor network contraction using~\cite{gray2021hyper},
and that of CutQC's brute force reconstruction.
The cost ratios are clipped at \textit{one billion} times improvement for clarity,
but the actual overhead reductions can be over $10^{50}\times$.
The classical post-processing cost ratios clearly demonstrate a fundamental improvement over~\cite{tang2021cutqc}.

\subsection{Quantum Area}
\begin{figure}[t]
    \centering
    \includegraphics[width=\linewidth]{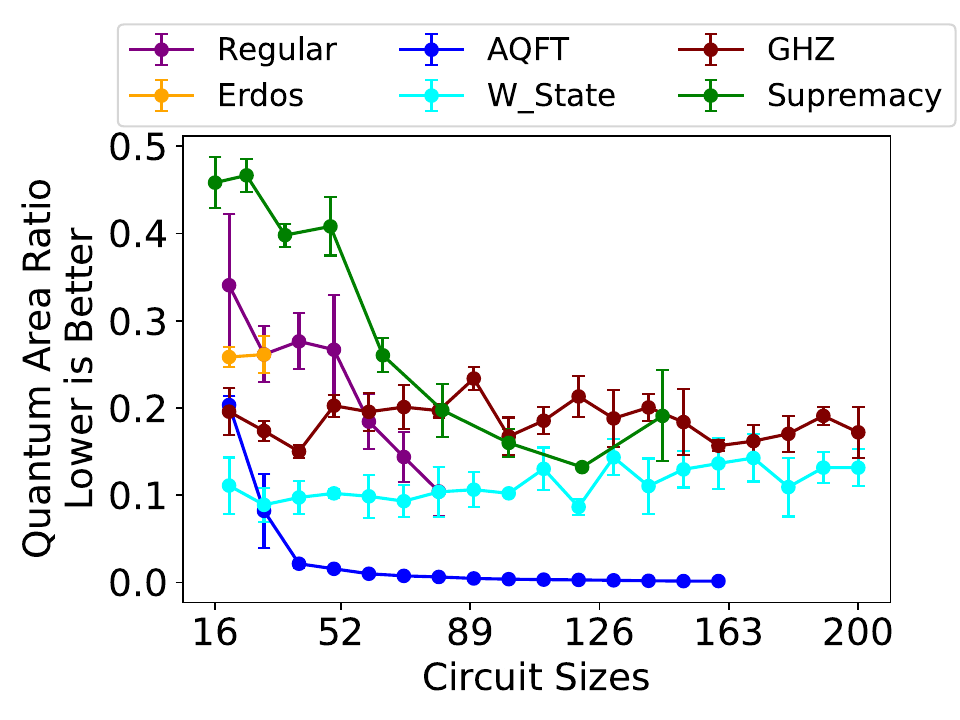}
    \caption{Quantum areas reductions of cutting versus not cutting for the same experiments in Figure~\ref{fig:total_runtime}.
    The plots show the quantum area of the largest subcircuit resulted from cutting as a percentage of the uncut benchmark.}
    \label{fig:quantum_area}
\end{figure}

Figure~\ref{fig:quantum_area} shows the quantum area reduction from cutting.
Quantum area for TensorQC is defined to be that of the largest subcircuit produced from cutting.
TensorQC only requires QPUs to support running the largest subcircuit at reasonable accuracies.
As a result, TensorQC reduces the QPU resource requirements by more than $10\times$ for various benchmarks.
QPU resource requirements reduction is at the expense of the extra cuts finding,
QPU executions of the subcircuits,
and tensor network contraction runtimes in Figure~\ref{fig:total_runtime}.

TensorQC opens up new potentials to run large scale quantum benchmarks using cutting and a set of less powerful QPUs.
Lower quantum area requirements translate to the ability to tolerate smaller and noisier QPUs.
Furthermore, in the fault-tolerant regime,
lower quantum areas translate to a looser requirement on the logical qubit error rate.
Depending on different quantum error correction solutions,
this implies much reduced physical qubit counts and error threshold requirements.
\subsection{HSS Data Efficiency}
\begin{figure}[t]
    \centering
    \includegraphics[width=\linewidth]{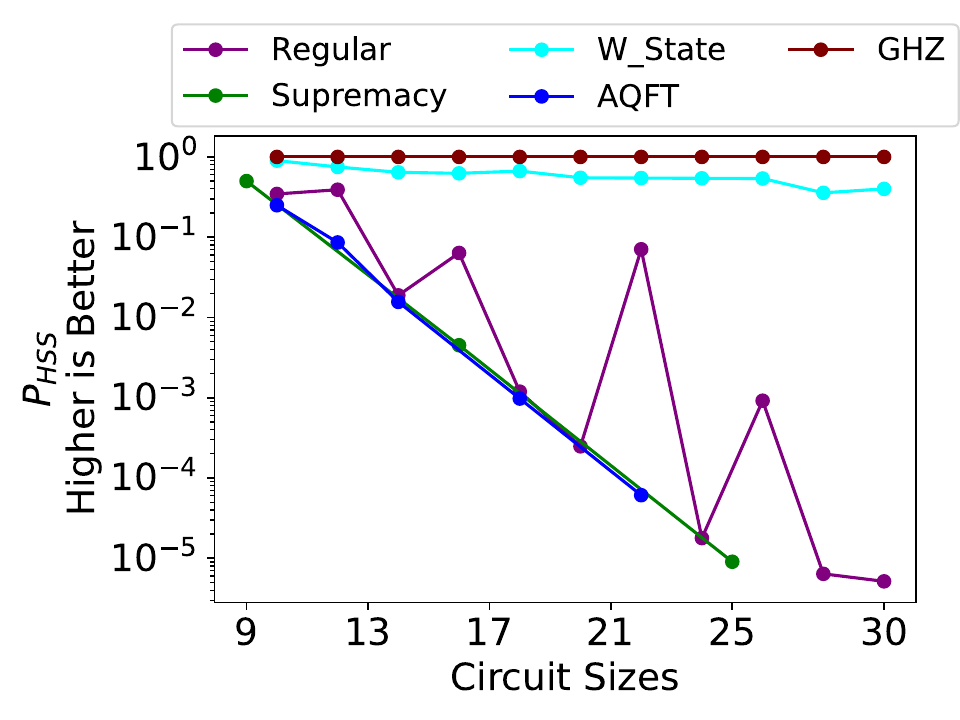}
    \caption{Retaining $2^8$ binary states using HSS for increasing benchmark sizes.
    HSS captures almost all the dominating states for benchmarks with skewed outputs
    including $W\_State$ and $GHZ$.
    Within our expectation,
    HSS captures decreasing amplitudes for other benchmarks with distributed outputs as sizes increase.
    }
    \label{fig:hss_p}
\end{figure}

\begin{figure}[t]
    \centering
    \includegraphics[width=\linewidth]{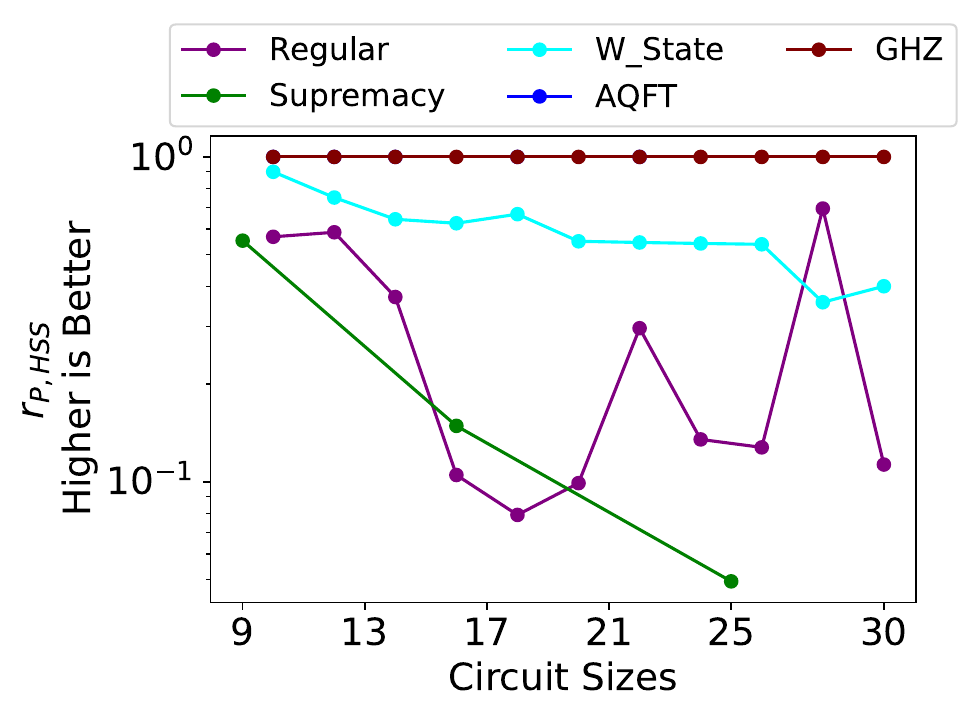}
    \caption{
    HSS amplitude retention as a percentage of the max possible retention with $2^8$ states.
    $AQFT$ (overlaps with $GHZ$ in the figure) shows nearly $r_{P,HSS}=1$ as any sampling algorithm performs the same for a uniform distribution.
    }
    \label{fig:hss_ratio}
\end{figure}

\begin{figure}[t]
    \centering
    \includegraphics[width=\linewidth]{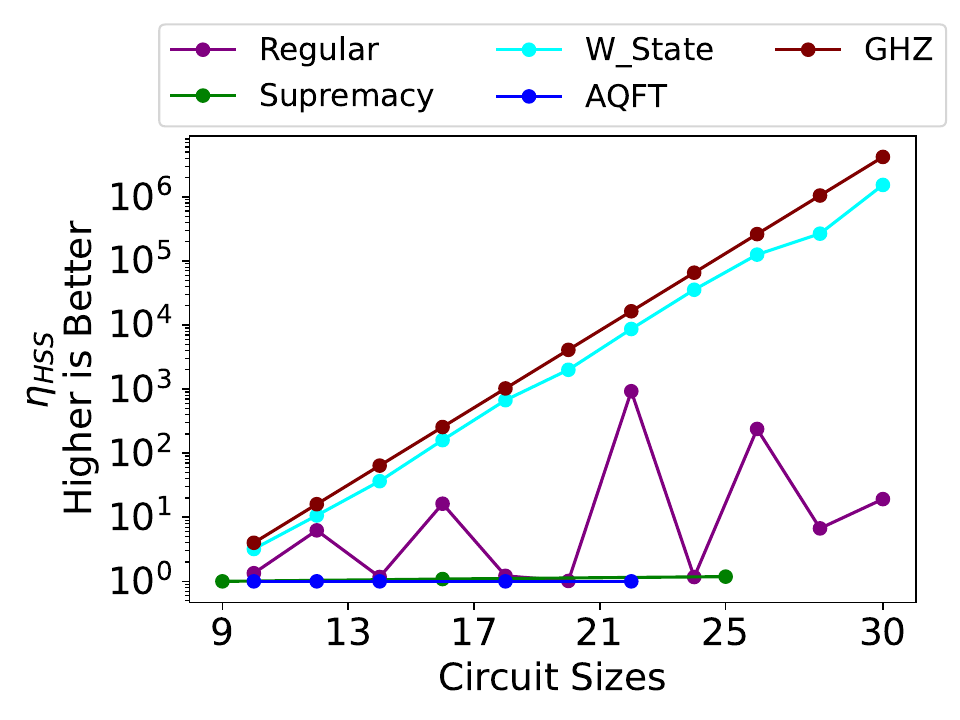}
    \caption{HSS data efficiencies for various benchmarks.
    }
    \label{fig:hss_efficiency}
\end{figure}

Reconstructing more states yields greater amplitude information
but also results in higher computational demands.
We limit the number of HSS reconstruction states to be $2^8$ in this section.
Figure~\ref{fig:hss_p} shows $P_{HSS}$ for various benchmarks with increasing sizes.

$W\_State$ and $GHZ$ benchmarks exhibit skewed binary state output landscapes,
allowing the HSS protocol to identify significant binary states with a minimal number of reconstruction states.
Notably, the HSS protocol retains nearly $100\%$ probability amplitudes at $30$-qubit while only reconstructing fewer than one-part-per-million binary states.

Benchmarks with more evenly distributed amplitude landscapes naturally pose a greater challenge for amplitude retention.
For example, $Regular$ benchmarks produce outputs without any dominating states.
Within our expectation,
the HSS algorithm retains a lower $P_{HSS}$ compared to benchmarks with skewed outputs.

However,
the lower amplitude retention is due to the intrinsic distribution of the benchmarks themselves.
Figure~\ref{fig:hss_ratio} shows that HSS captures a significant portion of the max possible amplitude retention,
despite the decreasing amplitudes for $AQFT$ and $Regular$ benchmarks shown in Figure~\ref{fig:hss_p}.

$\eta_{HSS}$ in Figure~\ref{fig:hss_efficiency}
shows that HSS retains much more amplitude weights relative to the few binary states reconstructed.
For example, HSS retains more than $10^6\times$ amplitudes than the number of states picked for $30$-qubit benchmarks with skewed outputs.

The $Supremacy$ benchmarks exhibit the poorest HSS metrics.
Unlike the $AQFT$ benchmarks,
which are uniform and thus allow any sampling algorithm to perform equally,
$Supremacy$ benchmarks feature a more evenly distributed output landscape.
This distribution complicates the task of identifying significant states.
Theoretically, a sampling algorithm could achieve a perfect $r_{P,HSS}=1$ across any output landscape,
but the diverse distribution in $Supremacy$ benchmarks makes this a particularly challenging goal.
\section{Related Work}\label{sec:related_work}
Many quantum compiler works exist to improve the performance of standalone QPUs to evaluate quantum circuits~\cite{nannicini2021optimal,tan2020optimality,murali2019noise,ding2020systematic,murali2020software,tang2024alpharouter}.
Quantum error correction is the key to build reliable QPUs~\cite{fowler2012surface,bravyi2012subsystem,javadi2017optimized,yoder2017surface,litinski2019game}.
QAOA uses classical computing to tune quantum circuit hyper-parameters to solve optimization problems~\cite{farhi2014quantum,tomesh2021coreset}.
However, they still rely entirely on QPUs to compute the quantum circuits.

Prior circuit cutting implementations~\cite{tang2021cutqc} rely on parallelization techniques for faster compute
while performing direct reconstruction of Equation~\ref{eq:reconstruction} with high overhead.
The various post-processing algorithms proposed in this paper go beyond what is possible from such techniques
and reduce the overhead itself.
Several small-scale demonstrations apply circuit cutting for chemical molecule simulations~\cite{eddins2022doubling},
variational quantum solvers~\cite{yuan2021quantum} and noise mitigation~\cite{basu2021qer}.

In addition, there are theoretical proposals to decompose a chemistry Hamiltonian on the algorithm level before translating it to a quantum circuit~\cite{eddins2022doubling}.
It is also theoretically possible to decompose $2$-qubit quantum gates~\cite{piveteau2023circuit},
instead of cutting quantum wires as in TensorQC.
There are proposals to manufacture distributed QPUs with quantum links~\cite{smith2022scaling,ang2024arquin}.
Quantum-linked QPUs essentially act as a large monolithic QPU and face even more hardware development challenges than building monolithic QPUs.
All these related works are completely orthogonal to TensorQC,
combining these techniques is a promising open question.
\section{Conclusion}\label{sec:conclusion}
This paper overcomes the post-processing scalability challenges for quantum circuit cutting
by incorporating tensor network contractions and developing heuristics cuts searching algorithms.
These techniques are necessary to demonstrate quantum circuit cutting for the large-scale benchmarks in the paper.
As a result, we demonstrate up to $200$-qubit benchmarks running on a single GPU,
which is the largest demonstration to the best of our knowledge.
Many benchmarks demonstrated in the paper are otherwise intractable on monolithic QPUs,
classical simulations, or even prior circuit cutting techniques.
Our hybrid workflow also reduces the quantum area requirement on QPUs by over $90\%$,
easing the demands on the QPU size and quality.
TensorQC hence bridges quantum computing with tensor networks and offers a practical way forward to running large-scale quantum benchmarks via cutting on distributed QPUs.
\section*{Acknowledgement}
This material is based upon work supported by the U.S. Department
of Energy, Office of Science, National Quantum Information Science Research Centers, Co-design Center for Quantum Advantage
(C2QA) under contract number DE-SC0012704.


\bibliographystyle{IEEEtranS}
\bibliography{refs}

\end{document}